\documentclass[aps,prb,superscriptaddress,twocolumn,no-footinbib,showpacs,floatfix,amsmath,amssymb]{revtex4-1}
 
\usepackage{graphicx}
\usepackage{dcolumn}
\usepackage{bm}
\usepackage[normalem]{ulem}
\usepackage{subfigure} 
\usepackage{soul,xcolor}
\usepackage{siunitx}
\usepackage{braket}
\usepackage{textcomp}
\setstcolor{red}
\RequirePackage{color}

\usepackage[capitalise,nameinlink,noabbrev]{cleveref}
\crefname{equation}{}{}
\crefname{enumi}{}{}

\begin{document}


\title{Topological origin of flat-bands as pseudo-Landau levels in uniaxial strained graphene nanoribbons and induced magnetic ordering  due to electron-electron interactions}
\author{Elias Andrade}
\affiliation{
 Posgrado en ciencias F\'isicas, Instituto de F\'{i}sica, Universidad Nacional Aut\'{o}noma de M\'{e}xico (UNAM). Apdo. Postal 20-364, 01000 M\'{e}xico D.F., M\'{e}xico
 }%
\author{Florentino L\'opez-Ur\'ias}
\affiliation{Divisi\'on de Materiales Avanzados, Instituto Potosino de Investigaci\'on Cient\'ifica y Tecnol\'ogica, Camino a la Presa San Jos\'e 2055, Col. Lomas 4a Secci\'on, San Luis Potos\'i, S.L.P 78216, M\'exico}
\author{ Gerardo G. Naumis}
\email{naumis@fisica.unam.mx}
\homepage{\\http://www.fisica.unam.mx/personales/naumis/}
\affiliation{
 Depto. de Sistemas Complejos, Instituto de F\'{i}sica, Universidad Nacional Aut\'{o}noma de M\'{e}xico (UNAM). Apdo. Postal 20-364, 01000 M\'{e}xico D.F., M\'{e}xico
 }%

\date{\today}

\begin{abstract}
Flat-bands play a central role in the presence of correlated phases in Moir\'e and other modulated two dimensional systems. In this work, flat-bands are shown to exist in uniaxially periodic strained graphene. Such strain should be produced for example by a substrate. The model is thus mapped into a one-dimensional effective Hamiltonian and this allows to find the conditions for having flat-bands, i.e., a long-wavelength modulation only on each one of the bipartite graphene sublattices,  while having a tagged strain field between neighboring carbon atoms. The origin of such flat-bands is thus tracked down to the existence of topological localized wavefunctions at domain walls separating different regions, each with a non-uniform Su-Schriffer-Hegger model (SSH) type of coupling. Thereafter,  the system is mapped into a continuum model allowing to explain the numerical results in terms of the Jackiw-Rebbi model and of pseudo-Landau levels. Finally,  the interplay between the obtained flat-bands and electron-electron interaction is explored through the Hubbard model.  The numerical results within the mean-field approximation indicate that the flat-bands induce N\'eel antiferromagnetic and ferromagnetic domains even for a very weak Hubbard interaction. The present model thus provides a simple platform to understand the physical origin of flat-bands, pseudo-Landau levels and the effects of the electron-electron interaction. 
\end{abstract}

\maketitle
\section{Introduction}
The study of Moir\'e superlattices has seen an explosion due to the experimental confirmation of correlation-driven electronic phases in twisted bilayer graphene (TBG), such as correlated insulating states\cite{cao2018correlated} and non-conventional superconductivity\cite{cao2018unconventional,yankowitz2019tuning}. Further research showed that other correlated phases can be found in twisted structures with more layers (twisted multilayers) \cite{PhysRevLett.123.026402,park2021tunable,park2022robust} , or heterostructures with different two-dimensional materials such as hexagonal boron nitride (hBN) \cite{PhysRevLett.122.016401,tran2019evidence,woods2021charge} or transition metal dichalcogenides (TMDs) \cite{naik2018ultraflatbands,devakul2021magic}. These twisted systems provide a platform for the study of correlated physics as the twist angle can tune the ratio between the strength of the interaction and the bandwidth. For certain angles known as magic angles flat-bands appear and the effects arising due interactions are enhanced, making possible the plethora of correlated phases found\cite{andrei2021marvels,sharpe2019emergent,cao2020tunable,zheng2020unconventional,cao2021nematicity}. Several studies have been made within the continuum model\cite{bistritzer2011moire,dos2012continuum,tarnopolsky2019origin,guinea2019continuum,carr2019exact,naumis2021reduction}, but the underlying mechanism of the unconventional superconductivity in TBG flat-bands is still under investigation. 

Flat-bands are not unique to Moir\'e materials, some lattices even have intrinsic flat-bands as result of the lattice geometry producing destructive interference\cite{liu2014exotic}. Another way to obtain flat-bands is through an external magnetic field \cite{goerbig2011electronic,tahir2020emergent}, as for strong enough fields it localizes the electrons in Landau orbits with a spectrum composed of flat Landau levels (LLs), however this breaks time reversal symmetry and requires extremely high magnetic fields. An alternative to this is strain \cite{levy2010strain,guinea2010energy,guinea2010generating,carrillo2014gaussian,NaumisReview,georgi2017tuning,andrade2019valley,liu2022analytic}, as it can induce pseudo-magnetic fields which have opposite action on each valley, such that time reversal is preserved and produce pseudo-Landau levels (pLLs) corresponding to fields with magnitudes of hundreds of Tesla. There are several experiments where the appearance of pseudo-magnetic fields have been observed \cite{PhysRevB.85.035422,jia2019programmable,PhysRevB.87.205405,nigge2019room,ma2018landau}, and recently more techniques to obtain flat-bands have been developed such as origami folding \cite{Yang2022} or buckled graphene\cite{mao2020evidence}. Recent studies have proposed models of periodically strained graphene with flat-bands that may reproduce some key aspects of TBG physics\cite{mele2020dirac,UntwistingMoire,NearlyFlatChernBands}. Particularly the flat-bands near the magic angles in TBG can be seen as zeroth pLLs originated from a pseudo-magnetic field generated by the Moir\'e pattern \cite{liu2019pseudo}. 

\begin{figure}[!htbp]
\begin{center}
{\includegraphics[width=.49\textwidth]{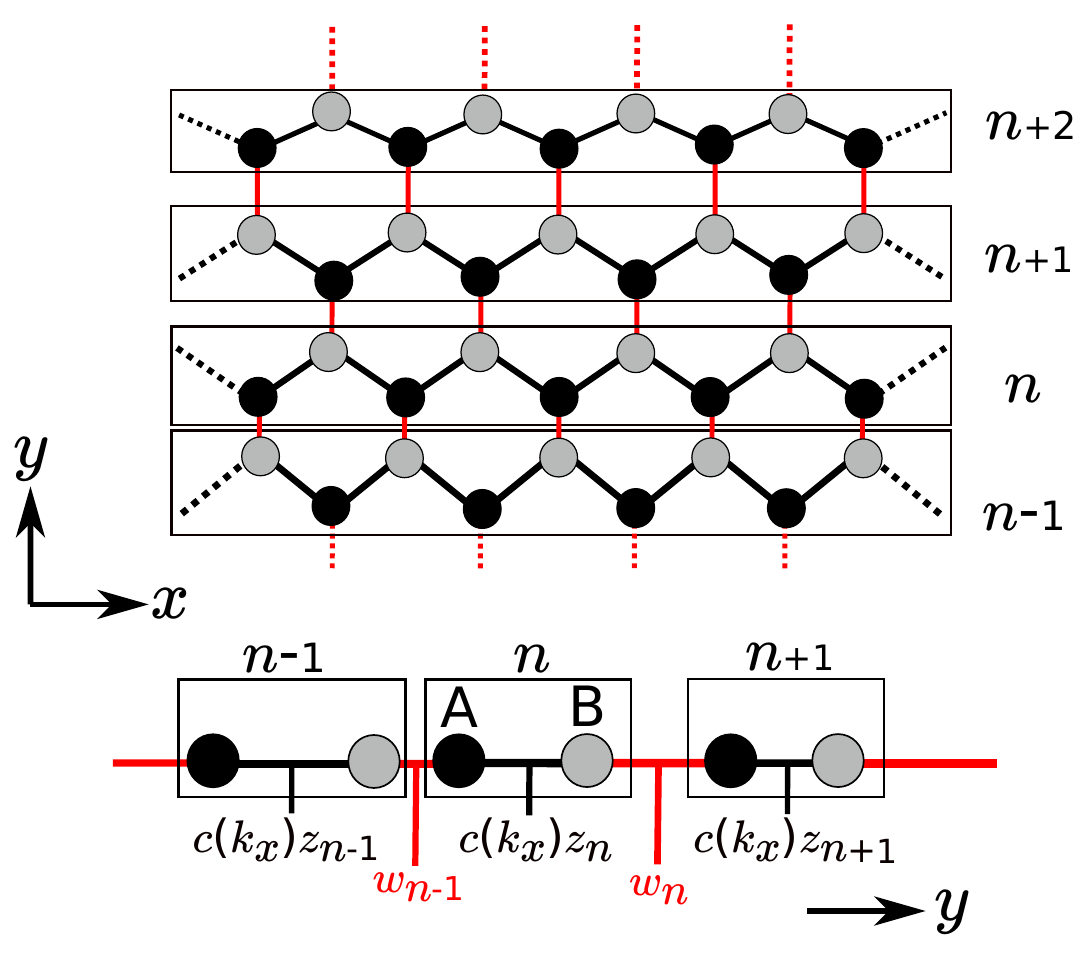}}
\caption{Graphene with a strain space dependent modulation along the $y$ axis with zigzag termination is mapped effectively into a 1D chain. Sites that belong to sublattice A and B are denoted by circles of color black and gray respectively.}
\label{fig:syst}
\end{center}
\end{figure}
Our interest is to analyze a model\cite{naumis2014mapping} containing flat-bands physics that has the advantage of being mapped to one dimension (1D), this model consists of the mapping of uniaxial strained graphene as shown in Fig. \ref{fig:syst}, the key ingredient to obtain flat-bands in this model is an oscillating strain with a wavelength slightly different from the distance between atoms of the same sublattice, the mismatch results in a Moir\'e wavelength several times larger than the original, but additionally the effective strain field has a phase difference between sublattices. This results in regions with different type of SSH\cite{su1979solitons} coupling separated through domain walls where the wavefunction is localized and the flat-bands can be regarded as soliton states\cite{Chamon2000,sasaki2010soliton}, for $E=0$ the regions around these domain walls become sublattice polarized and well separated between them. In the continuum limit we get the Jackiw-Rebbi model\cite{jackiw1976solitons} where the pseudo-magnetic field plays the role of an oscillating mass and the domain walls correspond to mass inversions in a one dimensional Dirac equation. To introduce the effects of correlations we add the Hubbard model to our Hamiltonian and solve it in the mean-field approximation, we show the appearance of N\'eel antiferromagnetic (AFM) and ferromagnetic (FM) domains.

The layout of this paper is as follows. In Sec. \ref{sec:Model}, we employ the 1D mapping of uniaxially strained graphene for different strain profiles, showing the appearance of flat-bands for oscillating strains with a wavelength slightly deviated from the sublattice periodicity. In Sec. \ref{sec:cont}, we derive a continuum model for the flat-band states, for $E=0$ we arrive to an analogue of the Jackiw-Rebbi model and compare it to our numerical results. In Sec. \ref{sec:pLL}, we derive a local Hamiltonian around the localization centers and obtain its spectrum as pseudo-Landau-Levels. In Sec. \ref{sec:e-e}, we introduce electron correlations through the Hubbard model and solve numerically within a mean-field approximation. Finally in Sec. \ref{sec:conclusions}, we discuss our results and present our conclusions.
\section{Model}
\label{sec:Model}
\begin{figure}[!htbp]
\begin{center}
{\includegraphics[width=.49\textwidth]{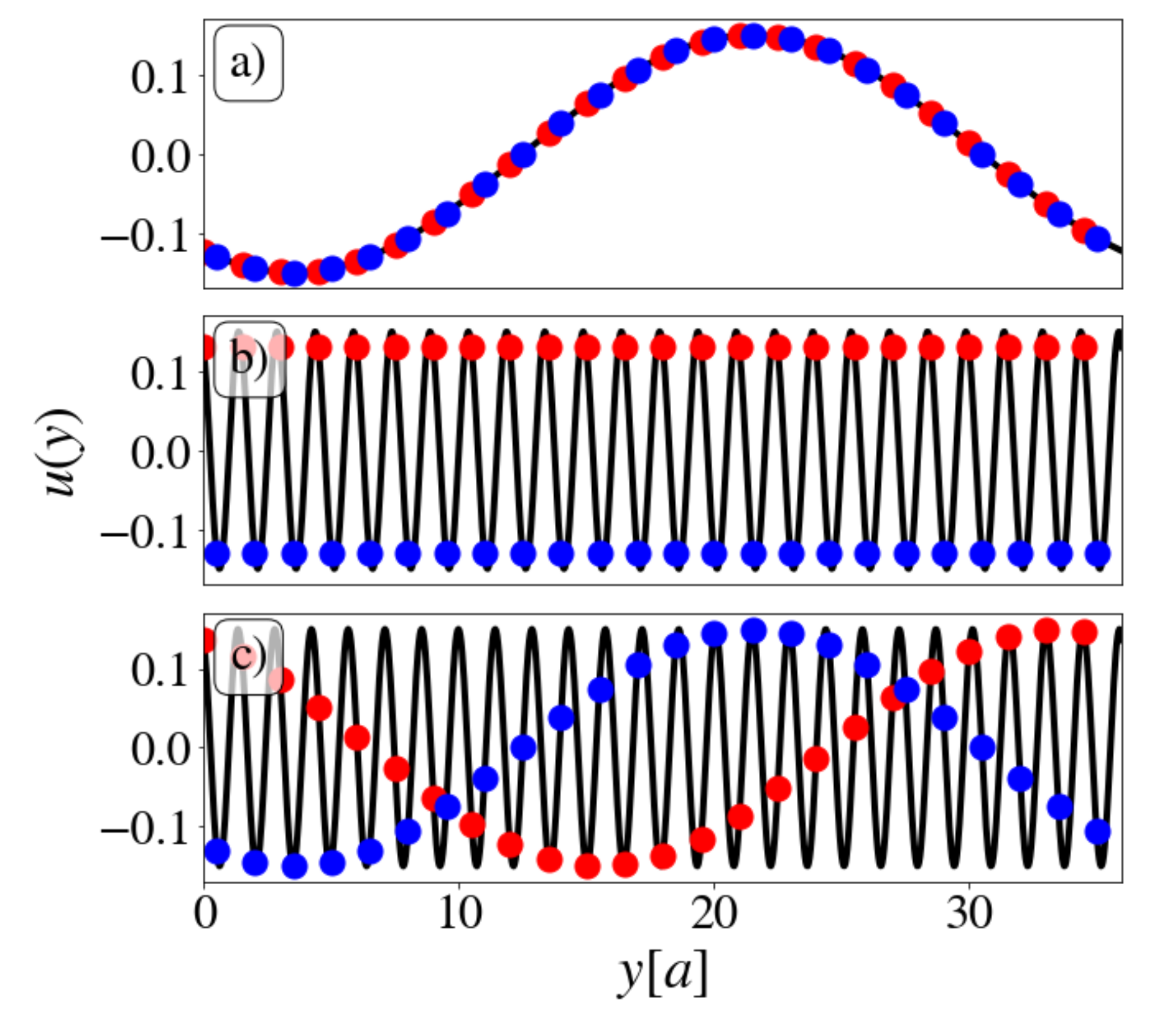}}
\caption{Displacement field $u(y)$ (black solid line). The values of the displacement field at the sites of sublattice A (B) correspond to the red (blue) dots. For a) $\lambda=36a$, b) $\lambda=\lambda_{sl}$ and c) $\lambda^{-1}=\lambda_{sl}^{-1}+(36a)^{-1}$. Here $\mu=0.15a$ and $\phi=5\pi/6$.}
\label{fig:displacement}
\end{center}
\end{figure}

We consider graphene with uniaxial strain along the $y$-direction, assuming a space dependent modulation $u(y)$ for a zigzag terminated nanoribbon, such that the atomic positions are changed as $(x',y')=(x,y+u(y))$. Considering the traslational symmetry along the $x$-direction, the system can be map into an effective 1D model as shown schematically in Fig. \ref{fig:syst}, this results in an effective Hamiltonian with $k_x$ dependent hopping elements\cite{naumis2014mapping},
\begin{equation}
    H(k_x)=-t_0\sum_n [c(k_x)z_n b_{n}^\dagger a_n+w_{n}a_{n+1}^\dagger b_{n}]+h.c.
\end{equation}
where $c(k_x)=2 \text{cos}(\sqrt{3}k_xa/2)$, $w_n$ and $z_n$ are the modulations of the hopping integrals which can be expressed in terms of the displacement field,
\begin{subequations}
\begin{equation}
    w_n=\text{exp}\left[ -\frac{\beta}{a}(u^A_{n+1}-u^B_n)\right],
\end{equation}
\begin{equation}
    z_n=\text{exp}\left[ -\frac{\beta}{2a} (u^B_{n}-u^A_n)\right],
\end{equation}
\end{subequations}
here $t_0 \approx 2.8$ eV is the hopping integral for pristine graphene, $\beta \approx 3$ is the Gruneisen parameter and $u^{A/B}_n$ is the value of the displacement at the $n$-th site of sublattice A/B, thus $u_n \equiv u(y_n)$. We consider an oscillating strain such as,
\begin{equation}
    u(y)=\mu \text{cos}\left(\frac{2 \pi}{\lambda} (y-a/2)+\phi\right),
    \label{Eq:u(y)}
\end{equation}
where $a \approx 1.42$ \AA $ $ is the distance between carbon atoms in pristine graphene, $\mu$ is the amplitude of the displacement, $\lambda$ is the wavelength of the oscillation and $\phi$ is an additional phase. For a wavelength greater than the lattice parameter $\lambda>>a$ the displacement field changes smoothly along the atomic positions as shown in Fig. \ref{fig:displacement} a). On the other hand if $\lambda$ is equal to the sublattice periodicity in the $y$-direction $\lambda_{sl}=3a/2$, each site of the same sublattice will see an equal displacement, this may result in a SSH Peierls distortion type of coupling along the $y$-direction, this is the case shown in Fig. \ref{fig:displacement} b).
Particularly if we consider a wavelength around $\lambda_{sl}$, such that:
\begin{equation}
    \frac{1}{\lambda}=\frac{1}{\lambda_{sl}}+\frac{1}{\lambda_{eff}}
\end{equation}
and substitute the positions of the atoms for each sublattice $y^A_n=n\lambda_{sl}$ and $y^B_n=n\lambda_{sl}+a/2$ in Eq. (\ref{Eq:u(y)}) we can get the effective displacement field for each sublattice:
\begin{subequations}
    \begin{equation}
        u^A(y)=\mu \text{cos}\left (\frac{2 \pi}{\lambda_{eff}}(y-a/2)-\frac{2\pi}{3}+\phi\right),
    \end{equation}
    \begin{equation}
        u^B(y)=\mu \text{cos}\left(\frac{2 \pi}{\lambda_{eff}}(y-a/2)+\phi\right),
    \end{equation}
\end{subequations}
the mismatch between the wavelength of the strain field oscillation and the sublattice periodicity produces a Moiré pattern with a longer effective wavelength $\lambda_{eff}$ but with a phase difference of $2\pi/3$ between both sublattices as shown in Fig. \ref{fig:displacement} c). 

\begin{figure}[!htbp]
\begin{center}
{\includegraphics[width=.49\textwidth]{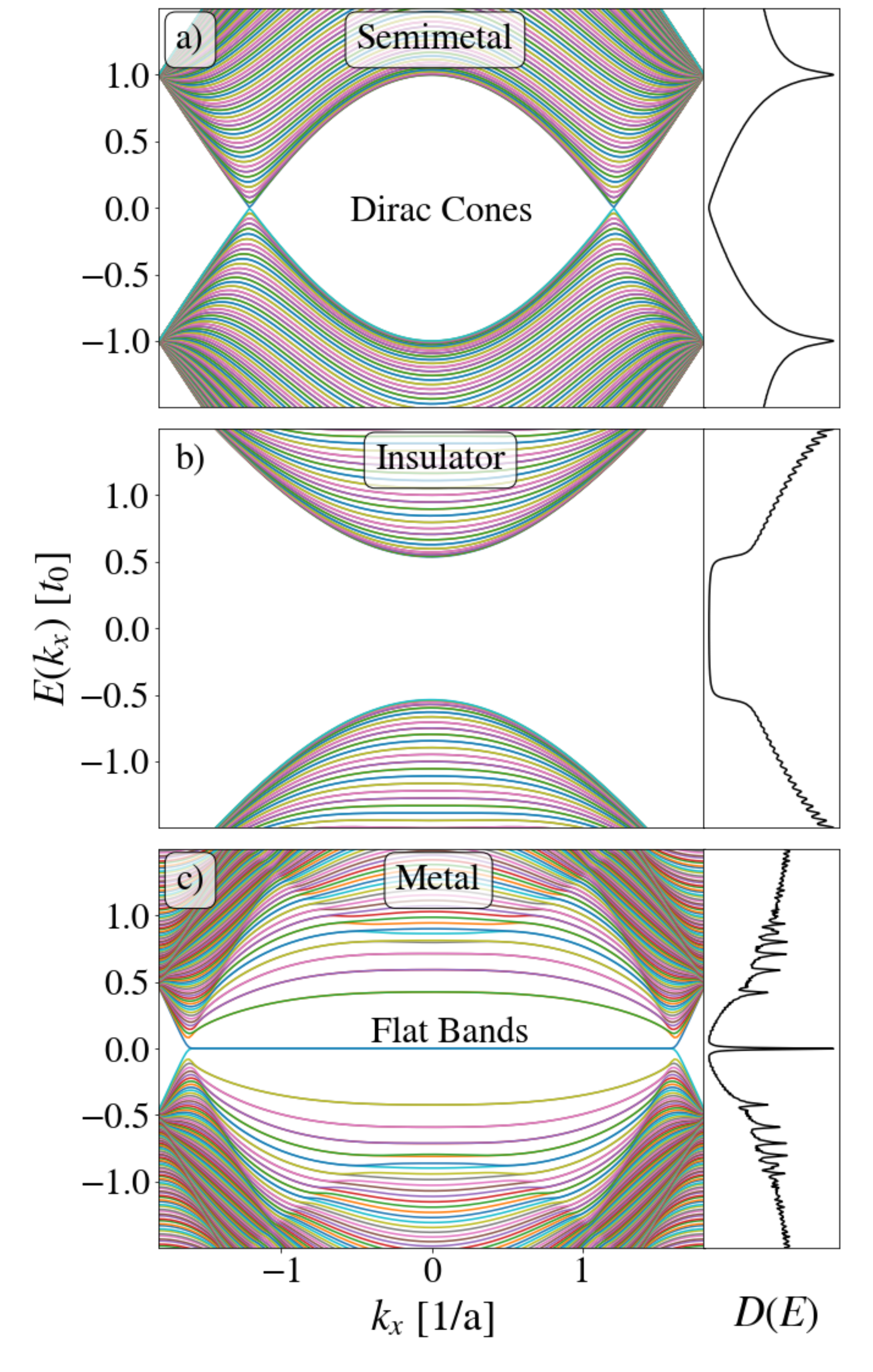}}
\caption{Left: Band structure for a) pristine graphene. b) Graphene under a strain field with $\lambda=\lambda_{sl}$ and $\mu=0.15a$ resulting in the opening of a gap. c) Graphene under a strain field with $\lambda_{eff}=240$ $a$ and $\mu=0.15a$, notice the clear appearance of flat-bands at $E=0$. Right: The corresponding density of states.}
\label{fig:bands}
\end{center}
\end{figure}
In Fig. \ref{fig:bands} we show the spectrum for different types of oscillations within the model, Fig. \ref{fig:bands} a) shows the dispersion for graphene without any strain, where the usual Dirac cones can be seen, in Fig. \ref{fig:bands} b) we consider an oscillation with $\lambda=\lambda_{sl}$ which opens a gap and in Fig. \ref{fig:bands} c) we consider $\lambda^{-1}=\lambda_{sl}^{-1}+(240\text{ }a)^{-1}$ which results in the appearance of flat-bands at $E=0$. This strain produces regions where one type of bond becomes shorter and the other longer and continuously change until their roles invert, the flat-bands arise due to soliton states at domain walls that separate this different regions, to see this consider the Schr\"odinger equation for the $n$-th atom of sublattice B,
\begin{equation}
    E\psi_n^B=-t_0[c(k_x)z_n\psi_n^A+w_n\psi_{n+1}^A],
\end{equation}
for zero energy modes the wavefunction becomes decoupled between sublattices as both have to satisfy the Schr\"odinger equation independently, thus we can obtain a recursion relation between neighboring atoms of the same sublattice,
\begin{equation}
    \psi_{n+1}^A=-c(k_x)\frac{z _n}{w_n}\psi_n^A,
\end{equation}
which can be rewritten as,
\begin{subequations}
\begin{equation}
    \psi_{n+1}^A=-c(k_x)e^{\frac{\beta}{a}(u_{n+1}^A+\frac{1}{2}u_n^A-\frac{3}{2}u_n^B)} \psi_n^A,
\end{equation}
and similarly for sublattice B,
\begin{equation}
    \psi_{n+1}^B=-\frac{1}{c(kx)}e^{\frac{\beta}{a}(u_{n}^B+\frac{1}{2}u_{n+1}^B-\frac{3}{2}u_{n+1}^A)}\psi_n^B,
\end{equation}
\end{subequations}
by applying these equations iteratively we can obtain the value of the wavefunction at any site given any initial value $\psi_0^{A/B}$, furthermore, since the displacement field changes slowly within the same sublattice we can consider $u_{n+1}^{A/B} \approx u_{n}^{A/B}$, which allows us to obtain a simpler expression, 
\begin{equation}
    \psi_n^{A/B}=[-c(k_x)]^{\pm n}\text{exp}\left[\pm \frac{3\beta}{2a}\sum_{j=0}^n \Delta u_j\right]\psi_0^{A/B},
    \label{Eq:DiscWF}
\end{equation}
where $\Delta u_j=u_j^A-u_j^B$. In the regions where $\Delta u_j$ is positive (negative) the wavefunction grows (decays) for sublattice A, while the opposite happens for sublattice B.
Thus the wavefunction for sublattice A is localized at the domain walls where $\Delta u_j=0$ going from positive to negative and for sublattice B where $\Delta u_j=0$ going from negative to positive.
\section{Continuum Model}
\label{sec:cont}
As the effective wavelength $\lambda_{eff}$ is greater than the sublattice periodicity $\lambda_{sl}$ it is feasible to consider a continuum limit where we use $y$ as a continuous variable. Thus we can consider our Hamiltonian as a $2\times2$ $y$-dependent matrix,
\begin{subequations}
\begin{equation}
    H(y)=
    \begin{pmatrix}
    0 & H_{AB}(y)\\
    H^*_{AB}(y) & 0
    \end{pmatrix},
\end{equation}
\begin{equation}
    H_{AB}(y)=-t_0[w(y)e^{ik\cdot\delta_1}+z(y)(e^{ik\cdot\delta_2}+e^{ik\cdot\delta_3})],
\end{equation}
where $\delta_j$ are the vectors connecting a site in the B sublattice to its three nearest neighbors in sublattice A, i.e., $\delta_1=a(1,0)$, $\delta_2=a(-\sqrt{3}/2,-1/2)$ and $\delta_3=a(\sqrt{3}/2,-1/2)$. Now we expand around the Dirac point $K'=\frac{4\pi}{3 \sqrt{3}a}(-1,0)$ of pristine graphene,
\begin{equation}
    H_{AB}(y)=A_x(y)-v_x(y)p_x+iv_y(y)p_y+\mathcal{O}(p^2),
\end{equation}
\end{subequations}
where,
\begin{subequations}
    \begin{equation}
        A_x(y)=t_0[z(y)-w(y)],
    \end{equation}
is a pseudo-magnetic potential arising due to the difference of the hopping amplitudes,
    \begin{equation}
        v_x(y)=v_fz(y),
    \end{equation}
and
    \begin{equation}
        v_y(y)=v_f \left(\frac{2}{3}w(y)+\frac{1}{3}z(y)\right),
    \end{equation}
are position dependent Fermi velocities, where $v_f$ is the usual Fermi velocity for pristine graphene defined as,
    \begin{equation}\maketitle
        v_f=\frac{3t_0a}{2 \hbar},
    \end{equation}
\end{subequations}
thus we can write our low-energy Hamiltonian as
\begin{equation}
    H(y)=(A_x(y)-v_x(y)p_x)\sigma_x-v_y(y)p_y\sigma_y
\end{equation}
where $\sigma$ are Pauli matrices acting on the sublattice pseudo-spin. Now we solve for the zero energy eigenstates $\Psi_0=(\psi^A_0,\psi^B_0)^T$,
\begin{equation}
    [(A_x(y)-v_x(y)p_x)\sigma_x-v_y(y)p_y\sigma_y]\Psi_0=0,
    \label{Eq:ContHam}
\end{equation}
since we have periodicity in the $x$ direction we consider it a good quantum number and substitute $p_x=\hbar q_x$, where $q_x$ is measured around the Dirac point. We then obtain the following Dirac equation for zero modes,
\begin{equation}
    [\partial_y\sigma_0-m(y,q_x)\sigma_z]\psi_0=0
\end{equation}
where we defined,
\begin{equation}
    m(y,q_x)=\frac{A_x(y)-\hbar v_x(y)q_x}{\hbar v_y(y)}
\end{equation}
we arrive to a continuum version of Eq. (\ref{Eq:DiscWF}),
\begin{equation}
    \psi_0^{A/B}(y,q_x)=N \text{exp}\left[\pm \int_y m(y',q_x)dy'\right],
    \label{Eq:wf}
\end{equation}
where N is a normalization constant. 
\begin{figure}[!htbp]
\begin{center}
{\includegraphics[width=.49\textwidth]{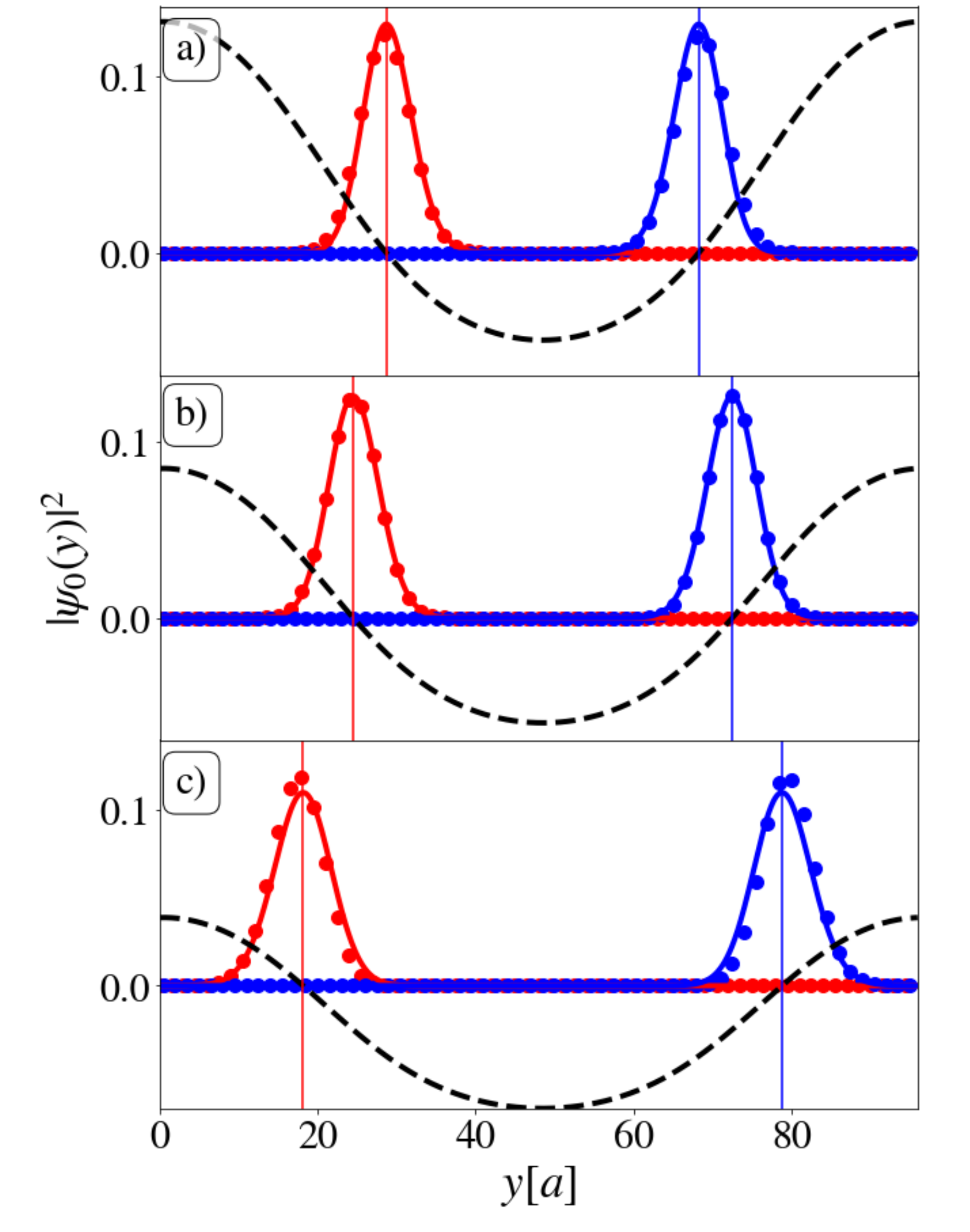}}
\caption{Probability density of the zero energy modes for a) $q_x=-0.15 a^{-1}$, b) $q_x=0$ and c) $q_x=0.15 a^{-1}$ with $\lambda_{eff}=96 a$ and $\mu=0.15a$, shown in red (blue) for sublattice A (B). The solid lines show the results obtained from the continuum limit in Eq. (\ref{Eq:wf}) while dots correspond to the results obtained through direct diagonalization of the system. The dashed black line shows $m(y,q_x)$, depending on whether it is positive or negative the density for one sublattice grows or decays. The localization centers are at the zeros of $m$ which are indicated by the thin vertical lines for each sublattice.}
\label{fig:WF}
\end{center}
\end{figure}

We can see $m(y,q_x)$ as a mass and our system becomes analogue of the Jackiw-Rebbi model\cite{jackiw1976solitons}, where a topological protected mode arises in the boundary between two regions with masses of different signs. In our case the mass oscillates along the $y$-direction resulting in the localization of the wave function around the zeros of $m(y,q_x)$, however the mass is seen with opposite sign between the two sublattices, thus in correspondence with the discrete case, the wavefunction of one sublattice is localized at the domain wall that changes sign from positive to negative and for the other sublattice in the opposite case. In Fig. \ref{fig:WF} we show the probability density for each sublattice, the solid lines show the solution obtained in Eq. (\ref{Eq:wf}) and the dots the solutions from direct diagonalization of the discrete system, notice the good agreement between both. The dashed black line shows $m$, due to its linear dependence on $q_x$ different values of $q_x$ will change the zeros of $m$, thus moving the localization centers.
\section{Pseudo-Landau Levels}
\label{sec:pLL}
\begin{figure*}
\centering
{\includegraphics[width=.95\textwidth]{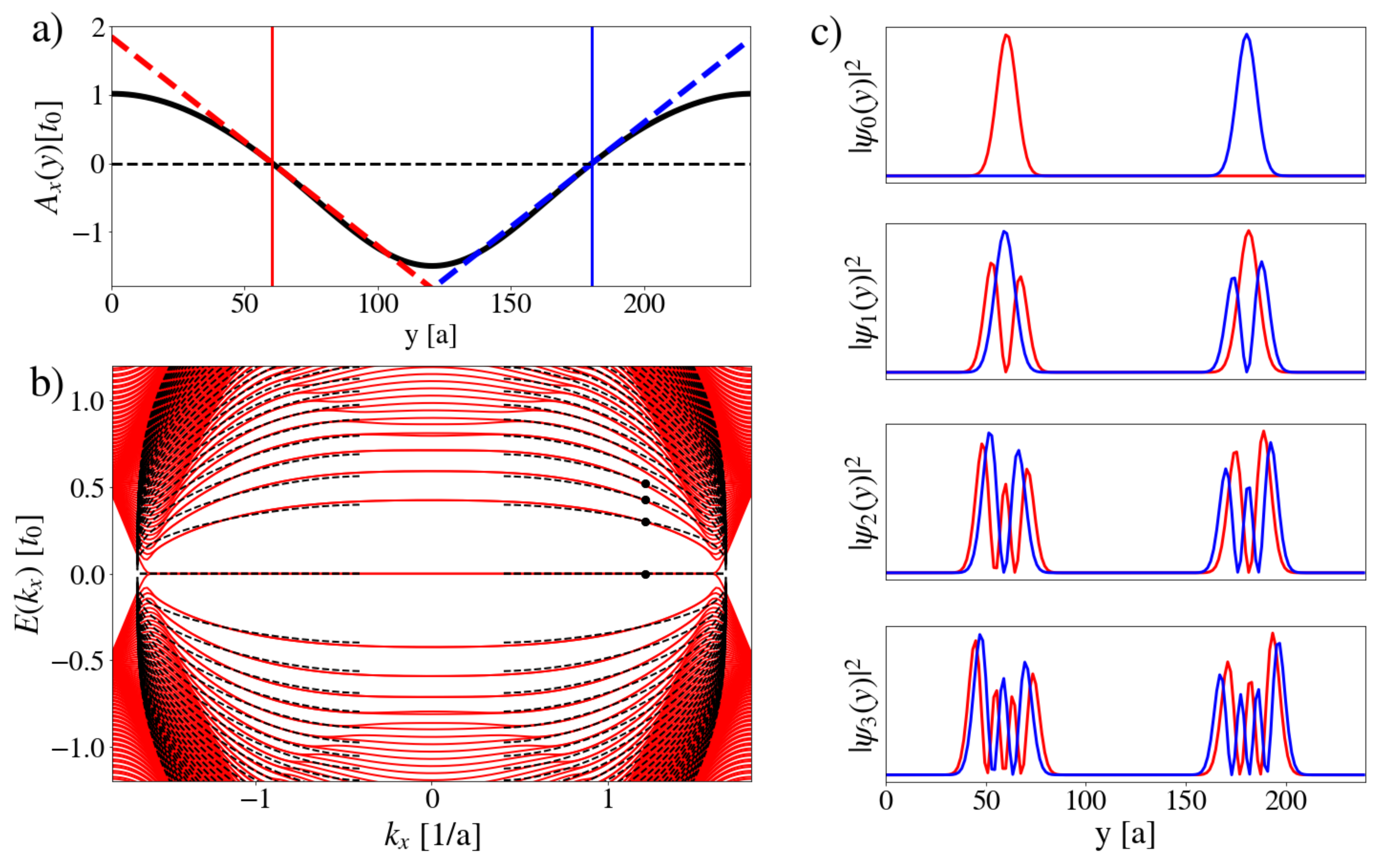}}
\caption{a) Pseudo-magnetic potential as function of $y$ (black solid line), for $q_x=0$ the positions at where the wavefunction is localized $y_0^{A/B}$ are shown by the vertical solid lines in red (blue) for sublattice A (B), around this points we perform an expansion to first order in $y$ as shown by the dashed lines. b) Pseudo-Landau levels (black dashed lines) and the band structure obtained from direct diagonalization (solid red lines). Notice the excellent agreement around $k_x=\pm1$. The localized states move closer between them as $k_x$ goes to zero, this results in the hybridization of pseudo-Landau levels localized at the two domain walls at higher energy states. At the boundaries of the Brilloun zone we have the dimer states with energies $E_n=\pm t_0 w_n$. c) The numerical probability density as function of $y$ for $n=0,1,2,3$ and $q_x=0$, these values are shown as black circles in b).}
\label{fig:pLL}
\end{figure*} 
As the wave function is localized around the points where $A_x(y)-v_x(y)\hbar q_x=0$, for a given $q_x$ we can calculate this positions as,
\begin{equation}
\begin{split}
    y_0^m(q_x)=&\frac{a}{2}+\frac{\lambda_{eff}}{2 \pi}\bigg[m\pi+ \frac{\pi}{3}-\phi \\
    &+(-1)^{m+1}\text{arcsin}\left(\frac{2a}{3 \sqrt{3} \mu \beta}\text{log}\left(1-\frac{3q_xa}{2}\right)\right)\bigg],
\end{split}
\end{equation}
the wavefunction is localized at the positions with odd $m$ for sublattice A and even $m$ for sublattice B. Furthermore, $A_x(y)$ is linear around $y_0^m$ as shown in Fig. \ref{fig:pLL} a) and the term  $\hbar v_x(y)q_x$ just produces a shift for small $q_x$, this allows us to expand up to first order in $y$, but for simplicity we expand only to zeroth order in $v_y(y)$, we then obtain the following local Hamiltonian,
\begin{subequations}
\begin{equation}
    h(y,q_x)=\omega(q_x)(y-y_0^m(q_x))\sigma_x+i\nu(q_x)\partial_y \sigma_y,
\end{equation}
where,
\begin{equation}
    \omega(q_x)=\partial_y[A_x(y)-\hbar v_x(y)q_x]\bigg\rvert_{y=y_0^m(q_x)},
\end{equation}
\begin{equation}
    \nu(q_x)=\hbar v_y(y_0^m(q_x)),
\end{equation}
\end{subequations}
from here onward we leave implicit the dependence of $q_x$. Now depending on whether the expansion is around a $y_0^m$ with an even or odd $m$ the slope of $A_x(y)$ changes in sign as shown in Fig. \ref{fig:pLL} a), resulting in two different local Hamiltonians $h^{\pm}$. We define the characteristic length as $l=\sqrt{\nu/\omega}$ and the dimensionless variable $\chi = (y-y_0)/l$, such that our Hamiltonians take the following form,
\begin{equation}
    h^{\pm}(\chi)=\sqrt{\nu \omega}
    \begin{bmatrix}
    0 & \pm \chi+\partial_\chi\\
    \pm \chi-\partial_\chi & 0
    \end{bmatrix},
\end{equation}
with the help of the annihilation $a=\frac{1}{\sqrt{2}}(\chi+\partial_\chi)$ and creation  $a^{\dagger}=\frac{1}{\sqrt{2}}(\chi-\partial_\chi)$ operators of the harmonic oscillator we can compactly write our Hamiltonians as,
\begin{equation}
    h^+= \epsilon
    \begin{bmatrix}
    0 & a\\
    a^{\dagger} & 0
    \end{bmatrix},
    \quad
    h^-=-\epsilon
    \begin{bmatrix}
    0 & a^{\dagger}\\
    a& 0
    \end{bmatrix},
\end{equation}
where $\epsilon=\sqrt{2\nu \omega}$. The spectrum is degenerate between both Hamiltonians and it is composed of pLLs,
\begin{subequations}
\begin{equation}
    E_n(q_x)=\pm\epsilon(q_x)\sqrt{n},
    \label{Eq:pLL}
\end{equation}
and the corresponding eigenstates for $n>0$ are given by,
\begin{equation}
\Psi^+_n(\chi)=\frac{e^{iq_xx}}{\sqrt{2}}
\begin{bmatrix}
\pm \psi_{n-1}(\chi)\\
\psi_{n}(\chi)
\end{bmatrix},
\end{equation}
\begin{equation}
\Psi^-_n(\chi)=\frac{e^{iq_xx}}{\sqrt{2}} \begin{bmatrix}
\psi_{n}(\chi)\\
\mp \psi_{n-1}(\chi)
\end{bmatrix},
\end{equation}
while for $n=0$,
\begin{equation}
    \Psi^+_0(\chi)=e^{iq_xx}
    \begin{bmatrix}
0\\
\psi_0(\chi)
\end{bmatrix},
\quad
    \Psi^-_0(\chi)=e^{iq_xx}
    \begin{bmatrix}
\psi_0(\chi)\\
0
\end{bmatrix},
\end{equation}
where $\psi_n$ are the eigenstates of the harmonic oscillator,
\begin{equation}
    \psi_n(\chi)=\left(\frac{1}{\pi 2^{2n}(n!)^2} \right)^{\frac{1}{4}}e^{-\chi^2/2}H_n(\chi),
\end{equation}
\end{subequations}
here $H_n$ are the Hermite polynomials.
In Fig. \ref{fig:pLL} b) we show the spectrum of the pseudo-Landau levels obtained in Eq. (\ref{Eq:pLL}) compared to the numerical band structure, they show an excellent agreement around $k_x=\pm1$. The additional structure around $k_x=0$ at higher energies comes from the hybridization between pseudo-Landau levels at opposite domain walls as they come closer for smaller values of $k_x$. At the borders of the Brillouin zone we have $c(k_x)=0$, thus dimer states appear with different hopping values along the $y$-direction, breaking the degeneracy, such that the spectrum is $E_n=\pm t_0 w_n$ in contrast to pristine graphene where it is simply $E=\pm t_0$. The numerical probability densities are shown in Fig. \ref{fig:pLL} c) for the first four states at $q_x=0$, here the structure composed of harmonic oscillator states with quantum number $n$ in one sublattice and $n-1$ in the other can be clearly seen. There is a slight asymmetry in the wavefunctions which is not present in our analytic results due to the approximations made, but the essential behavior is captured.

\section{Electron-electron interactions}
\label{sec:e-e}
As we are dealing with localized electrons with low kinetic energy, the contributions from electron-electron interactions become more relevant. Here we study these effects through the Hubbard model\cite{hubbard63,kanamori1963electron,gutzwiller1963effect,lieb1989two}, written in real space as:
\begin{equation}
H=\sum_{<i,j>,\sigma }t_{ij}\, \hat c_{i\sigma }^{\dagger} \hat c_{j\sigma
}+U\sum_{i} \hat n_{i\uparrow }\hat n_{i\downarrow }  
\label{eq:hubbard}
\end{equation}
{where $<i,j>$ denotes nearest-neighbor sites, $\hat c_{i\sigma }^{\dagger}$ ($
\hat c_{i\sigma }$) }refers to the creation (annihilation) operator for an
electron at site $i$ with spin $\sigma ${, $\hat n_{i\sigma }= \hat c_{i\sigma
}^{\dagger} \hat c_{i\sigma }$ }is the corresponding number operator and $%
t_{ij}=t_{ji} $ is the nearest-neighbor hopping integrals between the $i$th
and the $j$th sites. In this work, the parameter  $U$ is positive due 
to it is a direct Coulomb integral. Despite the simplicity of the model, the 
second term in Eq.~(\ref{eq:hubbard}) is not trivial from the computational 
point of view. This model can be solved exactly only for small systems since the
Hilbert space increases very rapidly with the number of sites. 
In this work, the Hubbard model is solved in the mean-field approximation. 
Thus, the second term in Eq.~(\ref{eq:hubbard}) is decoupled as 
\begin{equation}
U\sum_{i} (
n_{i\uparrow}   \langle \hat n_{i\downarrow} \rangle +  
n_{i\downarrow} \langle \hat n_{i\uparrow} \rangle  - 
\langle \hat n_{i\uparrow} \rangle \langle \hat n_{i\downarrow} \rangle  
) 
\label{eq:mfa}
\end{equation}
where $\langle n_{i\sigma} \rangle $ is the average electron occupation number with 
spin $\sigma$ at site $i$. A self-consistent solution is found iteratively  
by diagonalizing the Hamiltonian matrix over a uniform grid of $k$-points within the 
first-Brillouin zone. The iteration procedure is stopped when the changes of charge 
densities are less than  $10^{-6}$. The Fermi level is calculated from the integration 
of the total density of states (DOS). Then, $\langle \hat n_{i,\uparrow} \rangle$ and  
$\langle \hat n_{i,\downarrow} \rangle$ are obtained from the integration local DOS. 
The magnetic moment $\mu_{i}$ at the site $i$ is calculated as    
\begin{equation}
m_{i} = \frac{ \langle \hat n_{i\uparrow} \rangle - 
\langle \hat n_{i\downarrow} \rangle } {2} 
\label{mf2}
\end{equation}
The total magnetization is give by $M= \sum_{i} m_{i}$. The electronic charge 
$q_{i}$ at the site $i$ is given by 
\begin{equation}
q_{i} = \langle \hat n_{i\uparrow} \rangle + 
\langle \hat n_{i\downarrow} \rangle. 
\end{equation}
Thus, the total charge is given by $Q=\sum_{i} q_{i}$. In the half-filled band, $Q=N$
($N$ is the number of atoms). 


\begin{figure}[!htbp]
\begin{center}
{\includegraphics[width=.49\textwidth]{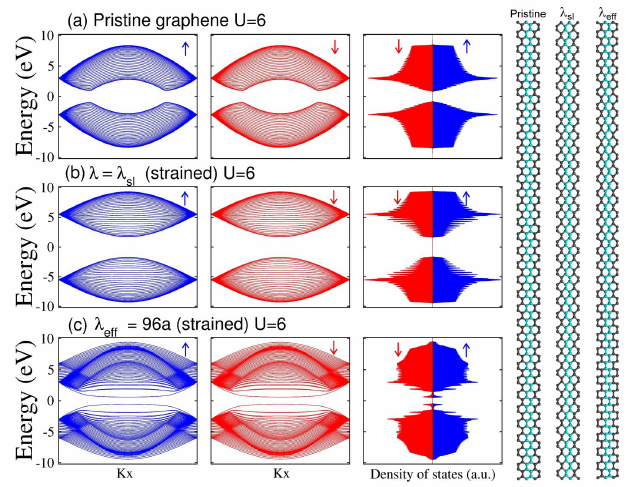}}
\caption{UHF-calculations of the band structure and total density of states for 
$U=6$ eV. (a) pristine graphene (first row panels), (b) strained graphene with 
$\lambda=\lambda_{sl}$ (second row panels), and (c) strained graphene with $\lambda_{eff}=9
6a$ (third row panels). The blue (red) color refers to states with spin up (down). 
The calculations correspond to $N=128$ atoms with one electron per atom (half-filled band)}
\label{fig:UHF}
\end{center}
\end{figure}
Fig.~\ref{fig:UHF} shows the spin-resolved band structure and density of states (DOS) for pristine 
and strained graphene considering $U=6$ eV. Band splitting, gap opening, and bandwidth narrow behavior 
are clearly promoted by the Coulomb repulsion. For pristine graphene, the band splitting 
occurs in all bands. The valence band maximum (VBM) and conduction band minimum (CBM) occur just in the Dirac 
point with a bandgap of 1.85 eV, see Fig.~\ref{fig:UHF}(a).  An inspection of the 
spin-resolved electronic charge in each site revealed that an antiferromagnetic ordering is developed for $U=6$ as 
shown later. Fig.~\ref{fig:UHF}(b) displays results for the strained graphene with $\lambda=\lambda_{sl}$. 
Even though the structure did not change the bandgap ($3.45$ eV) when compared with $U=0$, a narrow-band behavior 
was observed. This fact is related to the dimerization occurrence for $\lambda=\lambda_{sl}$, similar to the
SSH Peierls distortion. As shown above, the $\lambda=\lambda_{sl}$ case creates dimers with equal bond length and 
homogeneously distributed along the $y$-direction. The large and short bond lengths, resulting from the dimerization process, 
lead to charge confinement due to the kinetic-energy reduction in large bond lengths. This phenomenon occurs even in 
absence of the Coulomb repulsion as shown above. When the Coulomb repulsion is turned on ($U > \Delta$, where $\Delta$ is the bandgap)  in the dimerized system, the charge confined within the dimer is polarized and the energy is stabilized adopting an 
antiferromagnetic ordering, similar to a 
singlet state. Interesting electronic properties were also obtained for $\lambda_{eff}=96a$, see Fig.~\ref{fig:UHF}(c). 
Here the dimerization occurs, but in a non-homogeneous way along the $y$-direction, also a variation of 
the bond lengths exhibited changes, more details on it can be seen below. We observed a bandgap of $1.16$ eV with 
an extended flat-band behavior for VBM and CBM and a few neighboring bands around the $\Gamma$-point. These flat bands can be 
also appreciated as Van Hove singularities in the DOS. Notice that the flat-bands were also obtained for $U=0$ 
(two-fold degeneracy), but without the presence of a gap. The VBM has a two-fold degeneracy. The breaking of the 
two-fold degeneracy due to the Coulomb repulsion creates antiferromagnetic domains along the 
$y$-direction as we will show later.    
\begin{figure}[hbt!]
\centerline{\includegraphics[width=.45\textwidth]{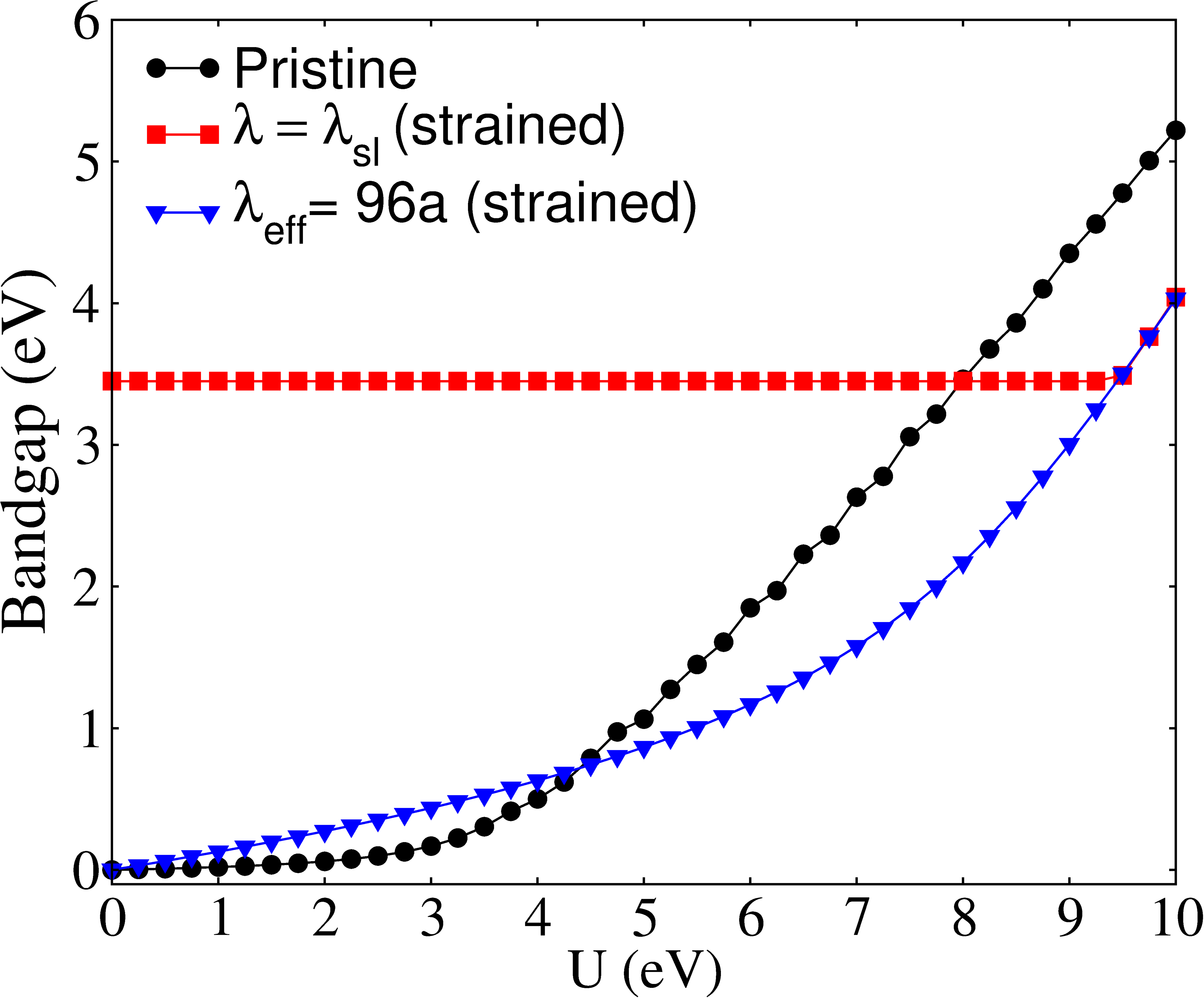}}
\caption{Bandgaps for pristine and strained graphene as a function of the on-site coulomb 
repulsion $U$. The systems contain $N=128$ atoms with one electron per atom (half-filled band). }
\label{fig:gaps}
\end{figure}
Different trends were observed in the bandgap as $U$ increased. In Fig.~\ref{fig:gaps}, we show the Coulomb 
repulsion dependence of the bandgap for pristine and strained graphene structures. We observed two 
linear behaviors separated by a critical Coulomb repulsion ($U_{c}$), which can be extracted 
from the inflection point of each curve. The pristine graphene showed $U_{c}\sim 4$, while strained 
graphene structures exhibited $U_{c}\sim 9.25$ and $6.5$ for $\lambda=\lambda_{sl}$ and  $\lambda_{eff}=96a$, respectively. 
Interestingly, in the strained graphene with $\lambda=\lambda_{sl}$, the bandgap remains unchanged for 
$U<9.2$. Furthermore, strained graphene with $\lambda_{eff}=96a$ exhibited a greater bandgap for $U< 4.4$ 
than the pristine graphene. 

\begin{figure}[hbt!]
\centerline{\includegraphics[width=.49\textwidth]{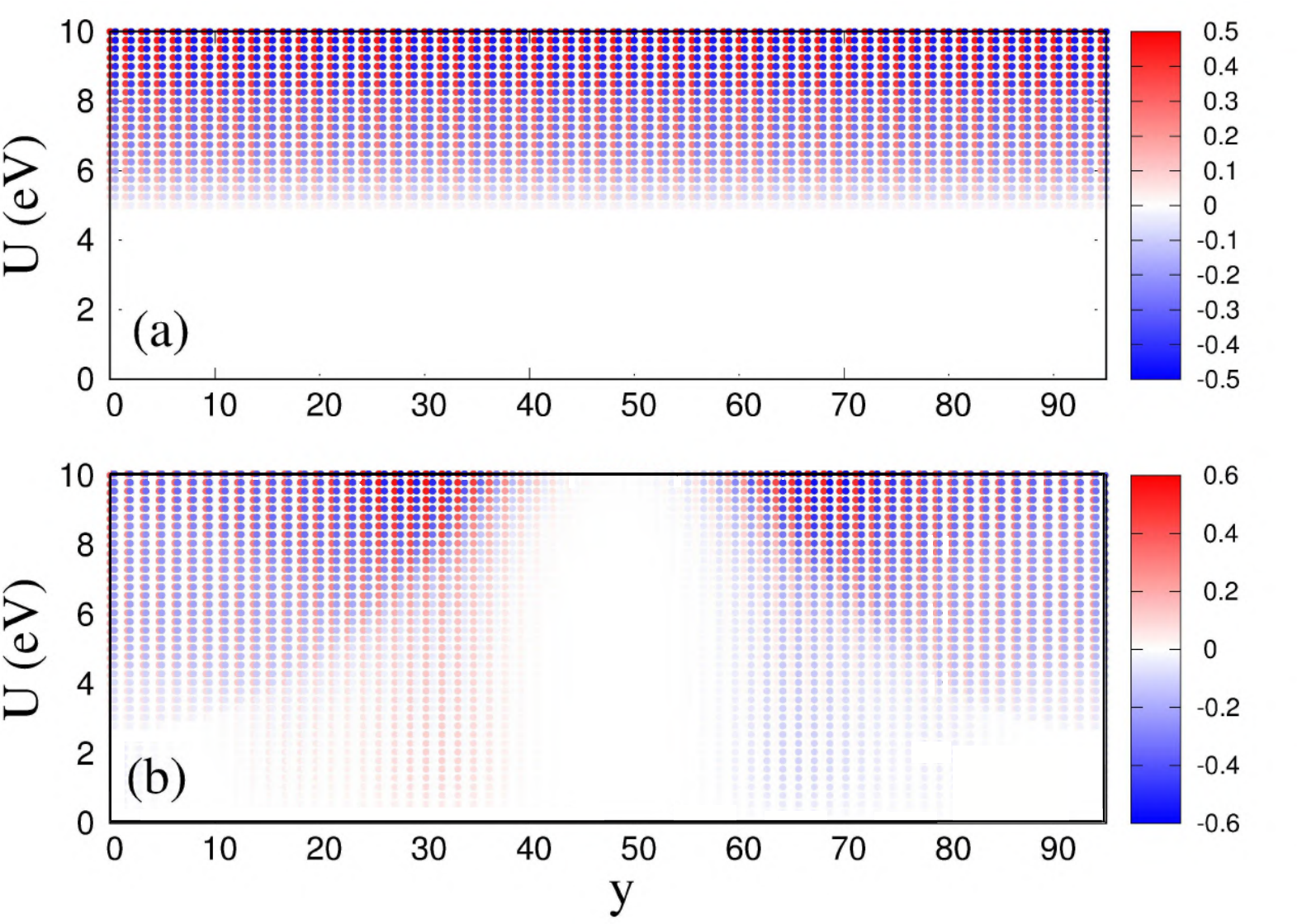}}
\caption{Magnetic ordering for (a) pristine graphene and (b) strained graphene with 
$\lambda_{eff}=96a$ for different values of on-site coulomb repulsion $U$. For pristine 
case, long-range N\'{e}el antiferromagnetic (AFM) order is developed for $U\ge 5$ eV as is indicated by
the alternated blue and red circle symbols along the y-coordinate. More intense color bar means an increment of 
the magnetic moments magnitude. For strained graphene (b), N\'{e}el AFM and FM 
domains are obtained. The calculations correspond to 
$N=128$ atoms with one electron per atom (half-filled band). Note that positive and negative 
magnetic moments are localized at sub-lattices A and B, respectively.}
\label{fig:mag_ord}
\end{figure}
Fig.~\ref{fig:mag_ord} shows the magnetic ordering evolution with $U$ along the $y$-direction. 
Results for pristine graphene clearly show two regions as $U$ increase, see Fig.~\ref{fig:mag_ord}(a). 
We observed paramagnetism for $U< 5$, and long-range AFM ordering for $U> 5$. The most intense colors (blue and red) 
refer to a strong localization regimen where local magnetic moments are close to $1/2$ (Heisenberg limit). More changes 
in the magnetic ordering along the $y$-direction can be seen for strained graphene with $\lambda_{eff}=96a$, as 
shown in Fig.~\ref{fig:mag_ord}(b). We observed different crossovers combining AFM, PM, and FM zones  
along the $y$-direction. For $U< 3$, the system exhibited FM domains identified as separated zones with the 
same color. For $U > 3$ separated AFM domains are obtained with a strong dependence on the local magnetic 
moments with the atom position along the $y$-direction.
\begin{figure}[hbt!]
\centerline{\includegraphics[width=.49\textwidth]{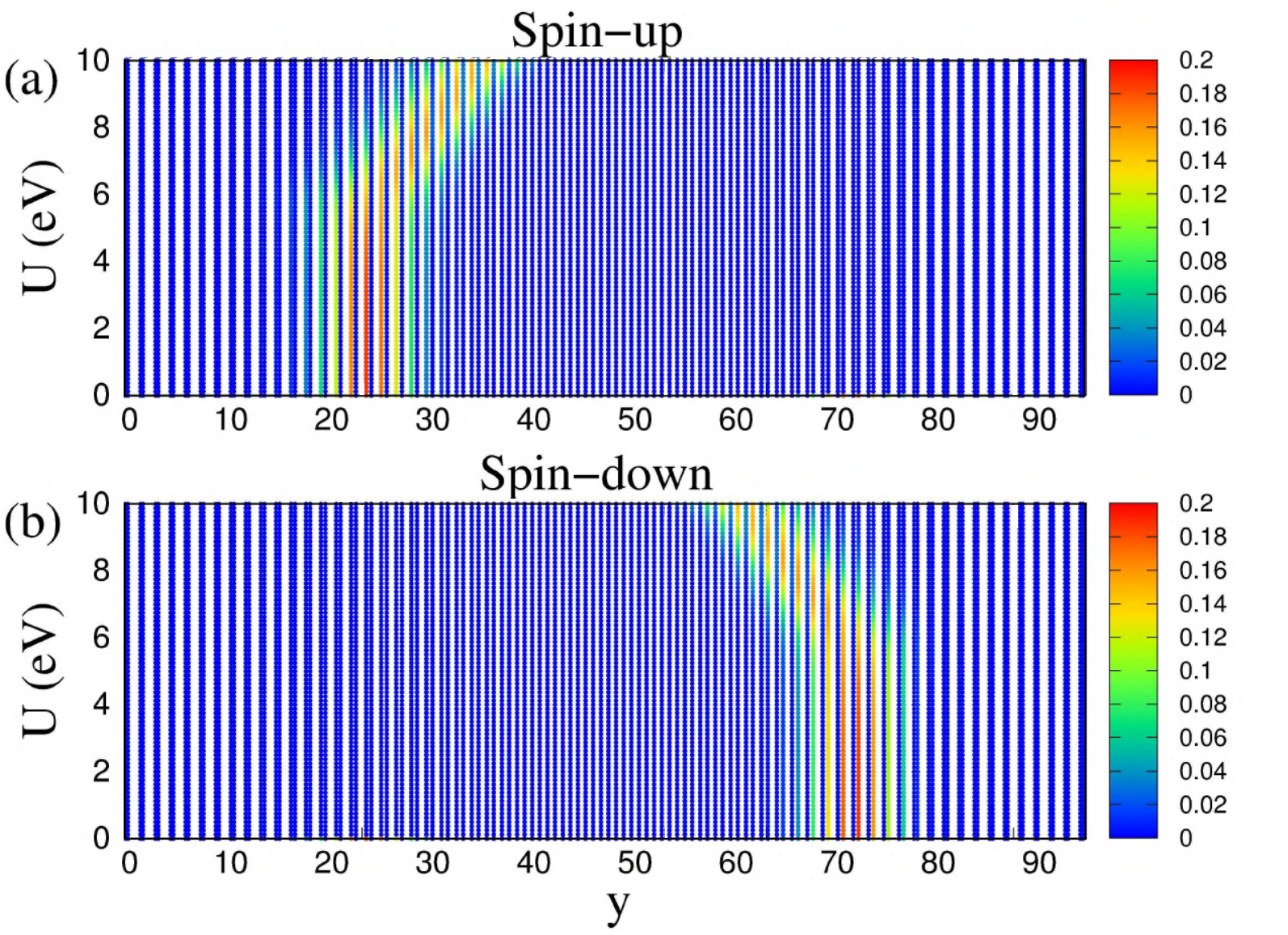}}
\caption{ 
Electron localization $|\Psi_{n} (K,y)|^{2}$ along the y-coordinate as a function of $U$ 
for strained graphene ($\lambda_{eff}=96a$), where $n$ corresponds to the valence band maximum (VBM)  
and  $K= 2\pi/ 3 \sqrt{3} $. The system corresponds to $N=128$ atoms with one electron 
per atom (half-filled band). Results for spin-up (a) and spin-down (b). Electrons with spin 
up and down are localized at sub-lattice A and B, respectively.}
\label{fig:res4}
\end{figure}
Fig.~\ref{fig:res4} displays the electron population in the VBM wave function at $K= 2\pi/ 3 \sqrt{3}$ for the strained 
graphene with $\lambda_{eff}=96a$. The blue (red) color means the null (maximal) probability of electron localization. The 
incorporation of the Coulomb repulsion makes that electrons with spin up and down live in two separate zones along the 
$y$-direction and different sublattices. Electrons with spin up and down live in A and B sublattices, respectively. 
For instance, in Fig.~\ref{fig:res4}(a), for spin-up electron localization around $y=24$, the colored vertical lines 
(sublattice-A) are accompanied on the right side by a blue vertical line (sublattice-B). This situation is reversed for 
spin-down electron localization around $y=72$, as shown in Fig.~\ref{fig:res4}(b). Note that the distance separation 
between spin-up and spin-down localization zones is reduced as $U$ increases. Both electrons with spin up and down exhibited 
maximal localization for $U< 4.5$, indicated by the red color. 
\section{Conclusions}
\label{sec:conclusions}
In this work we studied a 1-D model mapping of uniaxially strained graphene\cite{naumis2014mapping} and found the condition for the appearance of flat-bands as an effective displacement field that is out of phase between sublattices. These flat-bands can be described by solitons at domain walls and we provided analytical solutions in both discrete and continuum cases. In the continuum we obtained a connection to the Jackiw-Rebbi model and derived the pseudo Landau levels within a local approximation, the former corresponding to the zeroth pseudo Landau level. Electron-electron interactions were introduced by using a Hubbard Hamiltonian. The numerical results within a mean-field approximation indicate that flat-bands induce N\'eel antiferromagnetic and ferromagnetic domains. Also, the flat-band leads to electron spin polarization at different bipartite sublattices. Finally, the bandgap depends upon the long wavelength effective component of the strain, a fact that can be understood as a result of the electron-electron interaction effect in the charge confined within SSH dimers, where the energy is reduced by adopting an antiferromagnetic ordering.  
\section{Acknowledgments}
The authors acknowledge useful discussions with
Pedro Roman-Taboada and Andr\'es R. Botello-M\'endez. This work was supported by UNAM DGAPA PAPIIT IN102620 (E.A. and G.G.N.), CONACyT project 1564464 (E.A., F.L.-U.,G.G.N.). IPICYTs National Supercomputing Center supported this research with the computational time grant TKII-2021-FLU.

\bibliographystyle{apsrev}
\bibliography{refs.bib}
\end{document}